\documentclass[superscriptaddress]{revtex4}
\usepackage{graphicx}
\usepackage{amsmath}
\usepackage{floatflt,epsfig}
\def\ii{\'{\i}}
\def\d{\mbox{d}}

\begin{document}

\title{Quark stars within relativistic models}

\author{D.P. Menezes}
\affiliation{ Depto de F\ii sica - CFM - Universidade Federal de Santa
Catarina  Florian\'opolis - SC - CP. 476 - CEP 88.040 - 900 - Brazil}
\affiliation{School of Physics, University of Sydney, NSW 2006, Australia}
\author{C. Provid\^encia}
\affiliation{Centro de F\ii sica Te\'orica - Dep. de F\ii sica -
Universidade de Coimbra - P-3004 - 516 Coimbra - Portugal}
\author{D.B. Melrose}
\affiliation{School of Physics, University of Sydney, NSW 2006, Australia}

\begin{abstract}
Lately, it has been suggested that strange (quark) stars can be responsible
for glitches and other observational features of pulsars. Some discussions
on whether quark stars, if really exist, are bare or crusted are also a
source of controversy in the recent literature.
In the present work we use the Nambu-Jona-Lasinio model, known to
incorporate chiral symmetry, necessarily present in the QCD formalism, in order
to describe quark star properties. We compare our results for the stars
and the features of the model with the much simpler model normally used in the
description of strange stars, namely the MIT bag model. We also investigate the
differences in the stellar properties which arise due to the presence of the
crust. We show that the NJL model produces results which are somewhat different
as compared with the MIT model.
\end{abstract}

\maketitle

\vspace{0.5cm}
PACS number(s): 26.60.+c,24.10.Jv, 21.65.+f,95.30.Tg
\vspace{0.5cm}

\section{Introduction and formalism}

Neutron stars are the remnants of supernova explosions with masses 
1--$2M_\odot$, radii $\sim10\rm\,km$, and a temperature of the 
order of $10^{11}\rm\,K$ at birth, cooling within a few days to about 
$10^{10}\rm\,K$ by emitting neutrinos. In a conventional model for 
a neutron star, the star is composed of hadrons, predominantly 
degenerate neutrons with an admixture of protons, and degenerate 
electrons. In the stellar modeling, the structure of the star depends 
on the assumed equation of state (EOS), which contains a number 
of uncertainties associated with uncertainties in the strong force 
under conditions appropriate to a neutron star. An important 
uncertainty concerns the true ground star of matter.  In conventional 
models, hadrons are assumed to be the true ground state of the strong 
interaction. However, it  has been argued \cite{olinto,bodmer,witten} that 
`strange matter' is the true ground  state of all matter. If this is correct, 
the interior of neutron stars should be composed predominantly of 
$u,d,s$ quarks, plus leptons to ensure charge neutrality. This led to 
the suggestion that there may be no neutron
stars, and that all neutron-like stars are in fact strange stars 
\cite{olinto}. More generally, the composition of neutron stars 
remains a source of speculation, with some of the possibilities being 
the presence of hyperons \cite{aquino,magno,rafael}, a mixed phase 
of hyperons and quarks \cite{mp1,mp,pmp1,pmp2,trapping}, a phase of 
deconfined quarks or pion and kaon condensates \cite{kaons}. Models in which 
the interior is assumed to be composed of strange matter are often called 
`strange' stars. However, because, as we show, the strangeness 
content depends on the model used to describe the quark matter, we 
prefer to describe any model in which the interior involves 
deconvolved quarks (not bound in hyperons) as `quark' stars. In the 
stellar modeling, the structure of the star depends on the assumed 
EOS, which is different in each of these 
cases, and which depends on the nature of the strong interaction. 
Apart from the differences in the EOS, an important distinction 
between quark stars and conventional neutron stars is that the quark 
stars are self-bound by the strong interaction, whereas neutron stars 
are bound by gravity. This allows a quark star to rotate faster than 
would be possible for a neutron star. Further evidence in favor of 
quark stars is that some stars do indeed seem to rotate faster than 
what would be expected for a neutron star \cite{frieman,weber,Glen00}.

A quark star must have a thin layer on its surface dominated by the 
electrons which are necessary to enforce charge neutrality. This 
layer could suspend a hadronic crust, which would not be in contact
with the stellar core \cite{olinto,weber}. It has been suggested that 
the presence of a crust provides a natural explanation for pulsar 
glitches \cite{weber}, which are sudden changes in the rotation
period of the pulsar. Other authors have argued that quark stars 
should be `bare'  \cite{xureview,usov}, in the sense that any such 
crust would either not form or would be destroyed during the 
supernova explosion. Recently, the characteristics of the radiation 
from hot, bare strange stars have been identified \cite{usov04}.

Possible candidates for quark stars include compact star with 
gravitational masses around $\leq 1 M_{\odot}$ and radii of the order 
of $\simeq 6\rm\,km$:
PSR J1645-0317 (PSR B1642-03),
PSR J1830-1059 (PSR B1828-11) \cite{catalogue},
RX J185635-3754 \cite{pons02},
Her X-1 \cite{li95},
4U 1728-34 \cite{li99b},
X-ray bursters GRO J1744-28 \cite{cheng} and SAX J1808.4-3658 \cite{li99a}.

There are several different models that describe quark matter. In 
\cite{recentours} we investigated the properties of strange stars 
using two different models: the MIT bag model \cite{bag}, which is 
widely favored in the literature on strange stars, and the 
color-favor-locked phase (CFL) model \cite{cfl},
which allows the quarks near the Fermi surface to form Cooper pairs 
which condense and break the color gauge symmetry \cite{mga}. At 
sufficiently high density the favored phase is called CFL, in which
quarks of all three colors and all three flavors are allowed to pair. 
In \cite{recentours} we verified that quark stars have different 
maximum gravitational and baryonic masses, compared with neutron 
stars, and that the maximum masses are not significantly affected by 
the presence or absence of a crust. Moreover, with a simple 
prescription for the Kepler frequency, which determines the spin rate 
of a neutron-like star, we obtained much higher values for the quark 
stars than for the neutron stars. Finally, in
\cite{recentours} we used the two possible mass-to-radius ratio 
constraints available in the
literature \cite{cottam,sanwal} to distinguish between the assumed 
EOSs. Concerning one of the constraints, specifically the interpretation of 
absorption features as atomic transition lines in \cite{sanwal} is 
controversial: an alternative interpretation \cite{bignami,xu03} is 
that the aborption features are cyclotron lines, which imply no 
obvious constraint on the EOS. However, if the accepted 
interpretation is correct, then only the quark star EOS is compatible 
with the constraint given in \cite{cottam}, thereby excluding most of 
the neutron star EOSs. The conclusion that all neutron stars must be 
quark stars has major implications, but it is necessarily valid only 
for either of the two EOSs investigated in \cite{recentours}.

In the present paper we focus on a different model for the strange 
matter, the Nambu-Jona-Lasinio (NJL) model \cite{njl}. Our aim is to 
use this model to obtain the stellar properties and compare the 
results with the MIT bag model \cite{bag}. It is important to 
distinguish between the EOS during the short time period when 
neutrinos are still trapped in the star, and the EOS after the 
neutrons escape. The maximum entropy per baryon ($S$) reached in the core of a
new born star is about 2 (in units of Boltzmann's constant) \cite{prak97}.
We then perform our calculations for $S=0 (T=0$), 1 and 2 since the entropy
and not the temperature should be constant throughout the star \cite{burrows}.

In section II the formalism used is presented; in section III we give 
the results and make the relevant
discussions and in section IV the conclusions are drawn.

\section{Quark matter models}

In this section we summarize the main formulae for both models used 
in this paper.

\subsection{The Nambu-Jona-Lasinio Model}

We choose the NJL model \cite{Klevansky92,hk94,PRS87,Ruivo99} to 
describe the quark phase. The SU(3) version of the model includes 
most of symmetries of QCD, including chiral symmetry, and its 
breaking, which is essential in treating the lightest hadrons. The 
NJL model also includes a scalar-pseudoscalar interaction  and the 't 
Hooft six fermion interaction that  models the axial $U(1)_A$ 
symmetry breaking. The NJL model assumes deconfined point-like 
quarks, and is not remornalizable, requiring regularization through a 
cutoff in three-momentum space.

The NJL model is defined by the Lagrangian density
\begin{eqnarray}
L\,&=& \bar q\,(\,i\, {\gamma}^{\mu}\,\partial_\mu\,-\,m)\, q +\,g_S\,\,
\sum_{a=0}^8\, [\,{(\,\bar q\,\lambda^a\,
q\,)}^2\,\,+\,\,{(\,\bar q \,i\,\gamma_5\,\lambda^a\,
q\,)}^2\,]\nonumber\\
&+&\  \,g_D\,\,  \{{\mbox{det}\,[\bar q_i\,(1+\gamma_5)\,q_j]
+ \mbox{det}\,[\bar
q_i\,(1-\gamma_5)\,q _ j]\, }\},\label{1}
\end{eqnarray}
where $q=(u,d,s)$ are the quark fields and  $\lambda_a$ 
$(\,0\,\leq\,a\,\leq\,8\,)$ are the U(3) flavor
matrices. The model  parameters are:  $m\,=\, 
\mbox{diag}\,(m_u\,,m_d\,,m_s\,)$, the  current quark mass matrix 
($m_d=m_u$), the coupling constants $g_S$ and $g_D$ and the cutoff in 
three-momentum space, $\Lambda$.The NJL model is valid only for quark 
momenta smaller than the cut-off $\Lambda$.

The set of parameters is chosen in order to fit the values in vacuum 
for the pion mass, the pion decay constant,  the kaon mass and the 
quark condensates. We consider the set of parameters 
\cite{Ruivo99,kun89}:  $\Lambda=631.4\rm\,MeV$, $ 
g_S\,\Lambda^2=1.824$,  $g_D\,\Lambda^5=-9.4$,
$m_u=m_d=5.6\rm\,MeV$ and $m_s=135.6\rm\,MeV$  which are fitted to 
the following properties: $m_\pi=139\rm\,MeV$, $f_\pi=93.0\rm\,MeV$, 
$m_K=495.7\rm\,MeV$, $f_K=98.9\rm\,MeV$, $\langle\bar u 
u\rangle=\langle\bar d d\rangle=-(246.7\,\mbox{ MeV})^3$ and 
$\langle\bar s s\rangle=-(266.9\,\mbox{ MeV})^3$.

The   thermodynamical potential density is given by
${ \Omega}\, = \,{\cal E} -T\,S\,-\,\sum_{i} \mu_i N_i-\Omega_0$,
where the energy density is
  \begin{eqnarray}
{\cal E}\,& =&-2\, N_c \,\sum_i \int
{d^3 p\over (2\pi)^3}
{p^2 +  m_i M_i\over E_i} \,(n_{i-} - n_{i+})\,\theta (\Lambda^2 -p^2)\,
\nonumber\\
&-&
  2\,g_S\, \sum_{i=u,d,s}  \,\langle\,\bar q_i\,q_i\,\rangle^2
-2\,g_D\,\langle\,\bar u \,u\,\rangle\langle\,\bar d \,d\,\rangle\langle\,\bar
s \,s\,\rangle-{\cal E}_0\,\label{4}
\end{eqnarray}
and the entropy density is
\begin{eqnarray}
S=-2 N_c \sum_{i=u,d,s} \int {d^3p\over (2\pi)^3}\,\theta (\Lambda^2 -p^2)
\left\{[\,n_{i+} \mbox{ln}(n_{i+}) +
  (1-n_{i+})\, \mbox{ln}(1\,-\,n_{i+})]+ [n_{i+}\rightarrow n_{i-}]\right\}.
\label{5}
\end{eqnarray}
In the above expressions $N_c=3$, $T$ is the temperature,  $\mu_i$ 
($N_i$) is the chemical potential (number) of particles of type $i$, 
and ${\cal E}_0$ and $\Omega_0$ are included in order  to ensure 
${\cal E}= \Omega=0$ in the vacuum. This requirement fixes the 
density independent part of the EOS. The ground state of the system 
is described by the density  matrix \cite{Ruivo99} given by 
$f=\,\mbox{diag}\,(f_u\,,\,f_d\,,\,f_s)$ with
\begin{equation}
f_i = {1\over 2}\,[\,I\,(n_{i-}+n_{i+})\,-\,{  \gamma^0\,
M_i\,
+\,{\boldsymbol { \alpha}}\,\cdot\,{\bf p} \over E_i\,}
(n_{i-} - n_{i+})\,]\, \theta (\Lambda^2 -p^2),\label{2}
\end{equation}
where $I$ is the identity matrix, $n^{(\mp)}_i$ are the Fermi 
distribution  functions of the negative (positive) energy states, 
$n_i^{(\mp)} = [1+ \exp (\mp (\beta\,(E_i \pm \mu_i)))]^{-1}, \quad 
i=u,d,s.$
In the last equation  $\beta =1/T$, $M_i$ is  the constituent quark 
mass,  $E_i\,=\,(p^2\,+\,M^2_i)^{1/2}$.

The quark condensates and the quark densities are defined, for each 
of the flavors $i=u,d,s$,  respectively, as:
\begin{equation}
\langle\bar q_i\, q_i\rangle = -2 N_c\, \int {d^3 p\over (2\pi)^3}
{M_i\over E_i}  \,(\,n_{i-}\,-\,n_{i+}\,)\,\theta (\Lambda^2 -p^2), \label{6}
\end{equation}
\begin{equation}
\rho_i\,=\, \langle{q_i}^{\dagger}\, q_i\rangle = 2 N_c\,
\int {d^3p\over (2\pi)^3}\, (n_{i-} \,+\,
n_{i+}\,-1)\,\theta (\Lambda^2 -p^2).\label{7}
\end{equation}

Minimizing the thermodynamical potential $\Omega$ with respect to 
the constituent quark masses $M_i$ leads to three gap equations for 
the masses $M_i$
\begin{equation}
M_i\,=\,m_i\,-4\,g_S\,\langle\bar q_i\, q_i\rangle\,-\,2\,g_D\,\langle\bar
q_j\, q_j\rangle\langle\bar q_k\, q_k\rangle\,,\label{8}
\end{equation}
with cyclic permutations of  $i,\, j,\, k$.

We  introduce an effective dynamical Bag pressure \cite{bub99},
\begin{eqnarray*}
B&=& 2\, N_c \,\sum_{i=u,d,s} \int
{d^3 p\over (2\pi)^3}
\left(\sqrt{p^2 + M_i}-\sqrt{p^2 + m_{i}}\right)  \,\theta (\Lambda^2 -p^2)\,\\
&-&
  2\,g_S\, \sum_{i=u,d,s}  \,\langle\,\bar q_i\,q_i\,\rangle^2
-4\,g_D\,\langle\,\bar u \,u\,\rangle\langle\,\bar d \,d\,\rangle\langle\,
\bar s
\,s\,\rangle.
\end{eqnarray*}
In terms of this quantity the energy density (\ref{4}) takes the form
\begin{equation}
{\cal E}=2\, N_c \,\sum_{i=u,d,s} \int
{d^3 p\over (2\pi)^3}\sqrt{p^2 + M_i} \,\, (n_{i+}-n_{i-}+1)\, \,\theta
(\Lambda^2 -p^2)\,+B_{eff},\qquad B_{eff}=B_0-B,\label{ebageff}
\end{equation}
where $B_0=B_{\rho_u=\rho_d=\rho_s=0}$.  Writing the energy density 
in terms of $B_{eff}$ allows us to identify this contribution as a 
Bag pressure and establish a relation with the MIT Bag model 
\cite{njl,bub99} discussed in the next section.

\subsection{The MIT Bag Model}

The MIT Bag model \cite{bag} has been extensively used to describe quark 
matter. In its simplest form, the quarks are considered to be free 
inside a Bag and the thermodynamic properties are derived from the 
Fermi gas model. The energy density, the pressure and the quark $q$ 
density, respectively, are given by
\begin{equation}
{\cal E}= 3 \times 2  \sum_{q=u,d,s} \int \frac{\d^3p}{(2\pi)^3}
\sqrt{{\mathbf p}^2+ m_q^2} \left(f_{q+}+f_{q-}\right) + Bag,
\end{equation}
\begin{equation}
P =\frac{1}{\pi^2} \sum_{q}
\int \d p \frac{{\mathbf p}^4}{\sqrt{{\mathbf p}^2+m_q^2}}
\left(f_{q+} + f_{q-}\right) - Bag,
\end{equation}
\begin{equation}
\rho_q= 3 \times 2 \int\frac{\d^3p}{(2\pi)^3}(f_{q+}-f_{q-}),
\label{rhoq}
\end{equation}
where  $3$ stands for the number of colors, $2$ for the spin 
degeneracy, $m_q$ for the quark masses, $Bag$ represents the bag 
pressure and the distribution functions for the quarks and 
anti-quarks are the
Fermi distributions
  \begin{equation}
f_{q\pm}=1/({1+\exp[(\epsilon\mp\mu_q)/T]})\;,
\label{distf}
\end{equation}
with $\mu_q$ ($-\mu_q$) being the chemical potential for quarks 
(anti-quarks) of type $q$ and $\epsilon=({\mathbf p}^2+m_q^2)^{1/2}$. 
These equations apply at nonzero temperatures. For $T=0$, there are 
no antiparticles, and the particle distribution functions become the 
usual step functions.

We use $m_u=m_d=5.5\rm\,MeV$, $m_s=150.0\rm\,MeV$ and $Bag=(180{\,\rm 
MeV})^4$.  If $m_u,~m_d$ and $m_s$ are chosen as in the NJL model, 
the behavior of the properties of interest are not altered, since 
they are more dependent on the Bag pressure than on small differences 
in the quark masses.

\subsection{Quark matter in beta equilibrium}

In a star with quark matter we must impose both beta equilibrium and 
charge neutrality \cite{Glen00}. In what follows two scenarios are 
investigated, an early stage when there are trapped neutrinos in the 
interior of the star and a later stage, after the neutrinos escape 
(deleptonization). We first consider the later stage when entropy is 
maximum and neutrinos diffuse out. The neutrino chemical potential is 
then zero.  For $\beta$-equilibrium matter we must add the 
contribution of the leptons as free Fermi gases (electrons and muons) 
to the energy and pressure. The relations between the chemical 
potentials of the different particles are given by
\begin{equation}
\mu_s=\mu_d=\mu_u+\mu_e, \qquad  \mu_e=\mu_\mu.
\label{qch}
\end{equation}
For charge neutrality we must impose
$$\rho_e+\rho_\mu=\frac{1}{3}(2\rho_u-\rho_d-\rho_s).$$
For the electron and muon densities  we have
\begin{equation}
\rho_l=2 \int\frac{\d^3p}{(2\pi)^3}(f_{l+}-f_{l-}), \qquad
l=e,\mu,
\label{rhol}
\end{equation}
where  the distribution functions for the leptons are given in 
eq.~(\ref{distf}) by substituting  $q$ by $l$,
with $\mu_l$ the chemical potential for leptons of type $l$. At 
$T=0$, eq.~(\ref{rhol}) becomes
$\rho_l={k_{Fl}^3}/{3\pi^2}.$ The pressure for the leptons is
\begin{equation}
P_l= \frac{1}{3 \pi^2} \sum_l \int \frac{{\mathbf p}^4 dp}
{\sqrt{{\mathbf p}^2+m_l^2}} (f_{l+}+f_{l-}).
\label{pressl}
\end{equation}

In earlier stage, when the neutrinos are still trapped in the 
interior of the star, eq.~(\ref{qch}) is replaced
by
\begin{equation}
\mu_s=\mu_d=\mu_u+\mu_e -\mu_{\nu e},
\label{qchtrap}
\end{equation}
the lepton contribution is set to be $Y_L=Y_e+Y_{\nu e}=0.4$ 
\cite{burrows}. No muons appear in this case.

\section{Results}

In order to describe the crust of the quark stars we next use the 
well known EOS calculated in \cite{bps} for very low densities.

\begin{figure}
\includegraphics[width=10.cm]{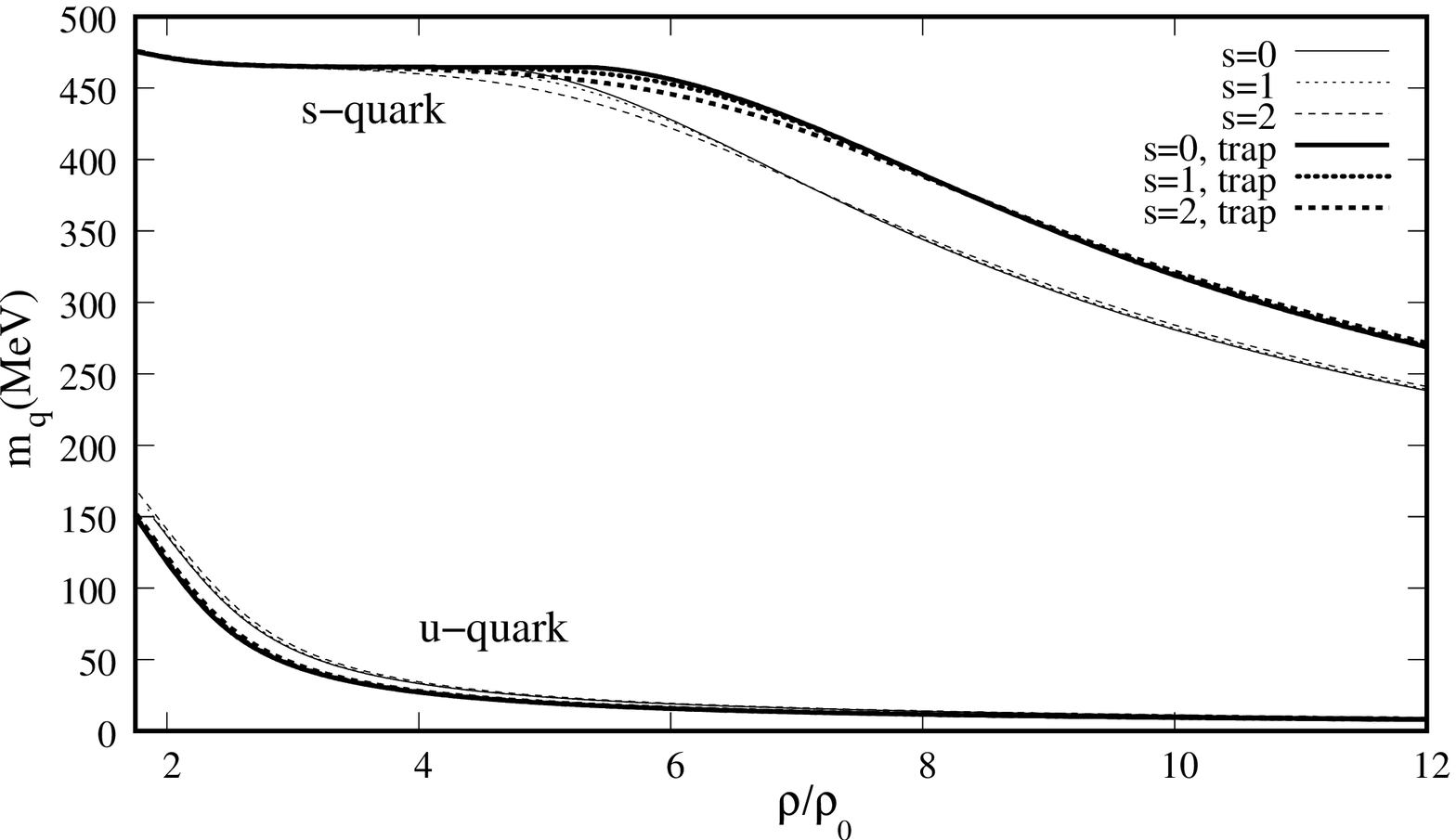}
\caption{the $u$ and $s$-quark mass as a function of density within the NJL  
for the different entropies with (thick lines) and without (thin lines) 
neutrino trapping}
\label{mu}
\end{figure}

In figure \ref{mu} the strange quark mass as a function of the 
density is shown for the NJL models and the 6 cases discussed in the 
present work, i.e., $S=0,1,2$ without neutrinos and with trapped 
neutrinos.
These data have a substantial influence on the results we obtain with 
the NJL model. Chiral  symmetry restoration occurs first for the $u$ 
and $d$ quarks and only at higher densities for the $s$ quarks. This 
is clearly seen in figure \ref{mu}: the mass of the $s$-quark starts 
to decrease only for $\rho/\rho_0>4\, (6)$ for neutrino free 
(trapped) matter.

\begin{figure}
\begin{tabular}{cc}
\includegraphics[width=10.cm]{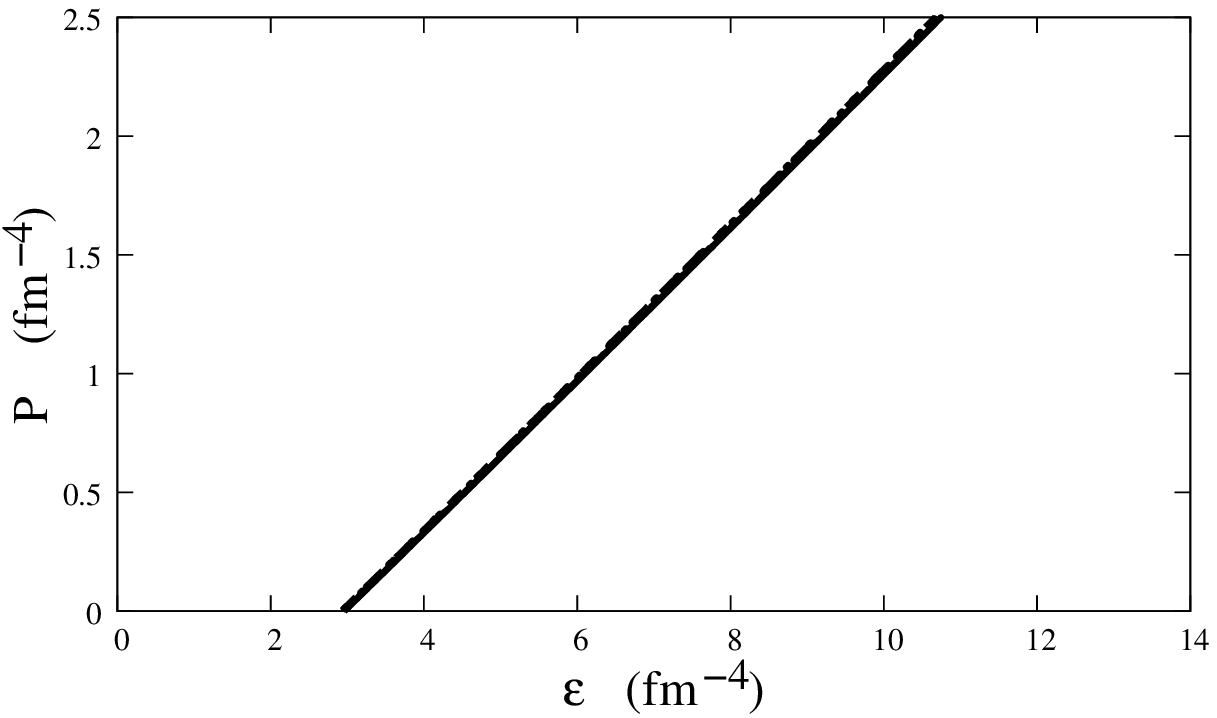}\\
\includegraphics[width=10.cm]{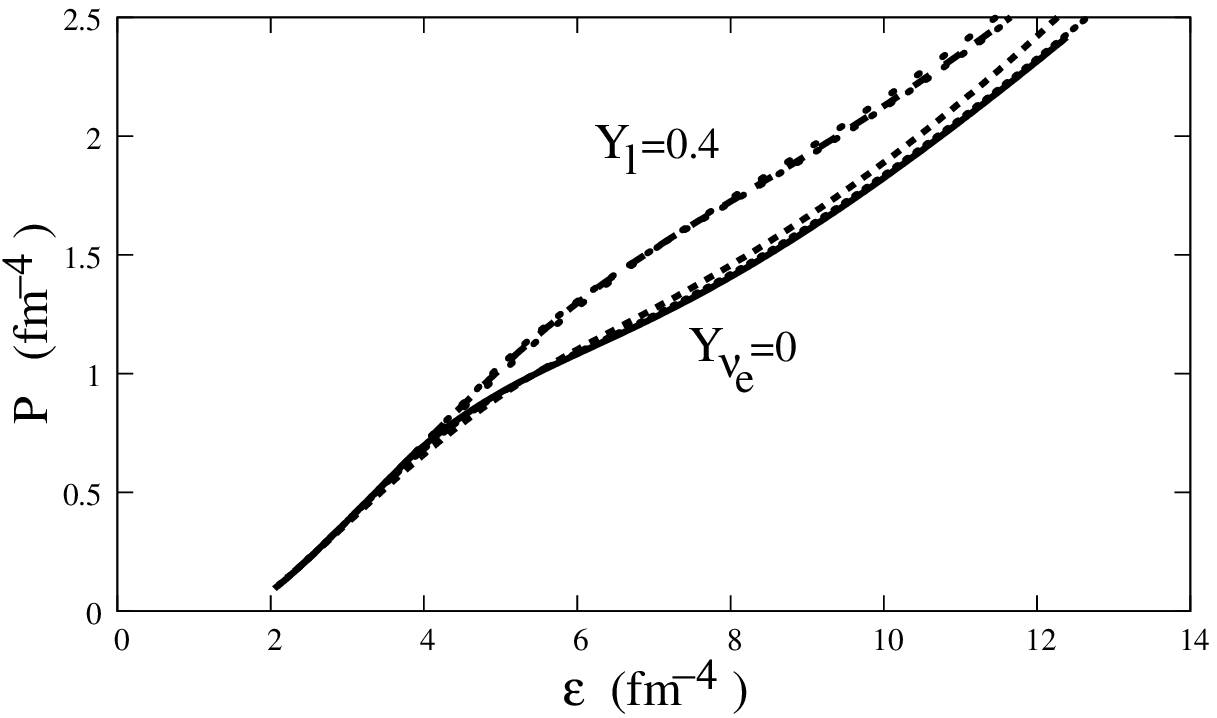}\\
\end{tabular}
\caption{EOS obtained for $S=0,1,2$ without neutrinos and with trapped 
neutrinos within the MIT bag model (top figure) and the NJL model 
(bottom figure). $S=0$ gives always the softer and $S=2$ the harder EOS.}
\label{fig1}
\end{figure}

In figure \ref{fig1} we show the EOS obtained respectively for the 
MIT bag  and for the NJL models again for the 6 different cases 
mentioned above. As shown in the upper panel in figure \ref{fig1},
the results are very similar and it is difficult to distinghish 
between them; this may be attributed to the familiar 
result that for an extreme relativistic gas the pressure is one third of the 
energy density. The NJL model presents different behaviors at high 
densities. The fact that chiral symmetry is
restored at different densities for quarks with different masses 
explains the softer behavior of the EOS for $\epsilon> 
4\rm\,fm^{-4}$. At these densities the onset of  strangeness occurs, 
with the onset occurring at larger densities when neutrino trapping 
is included (see Fig.~\ref{mu}). The EOS with trapped neutrinos is 
harder and the $S=0,1,2$ cases are similar, but distinghishable since 
higher temperatures correspond to harder EOSs. This behavior was 
already seen in \cite{trapping}, although in a different context.

\begin{figure}
\begin{tabular}{cc}
\includegraphics[width=10.cm]{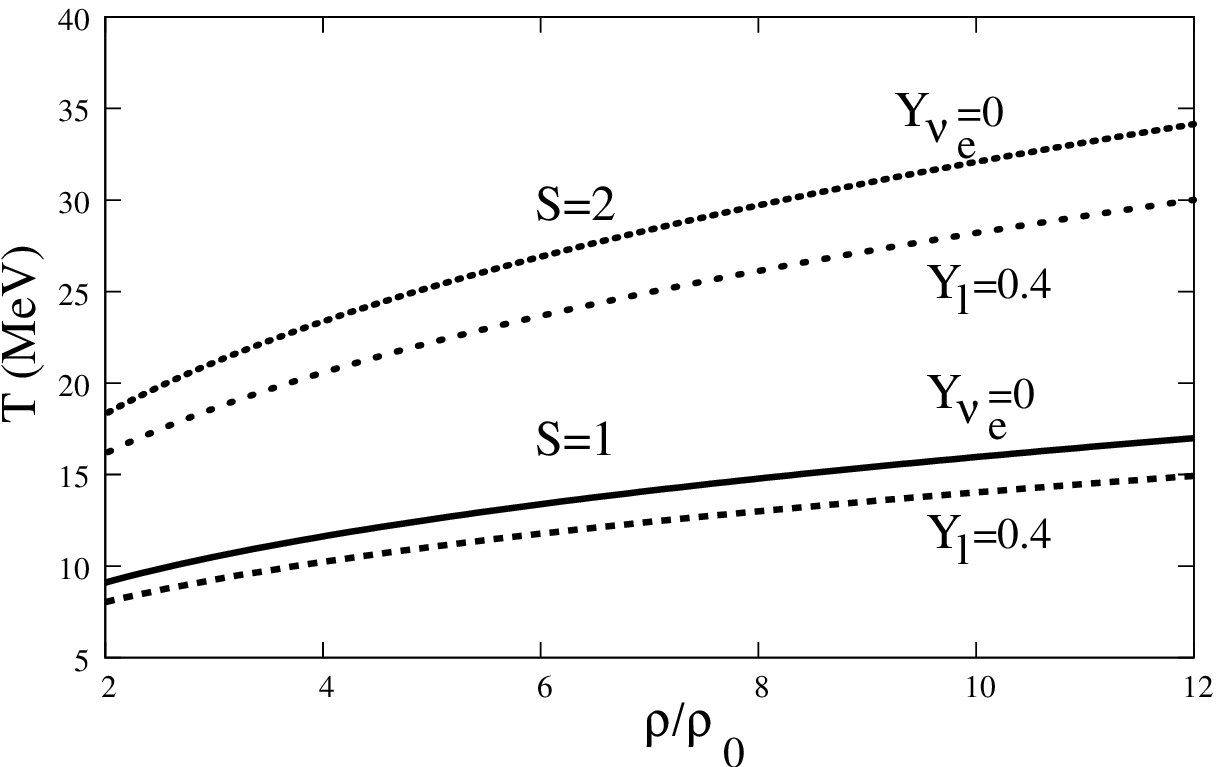}\\
\includegraphics[width=10.cm]{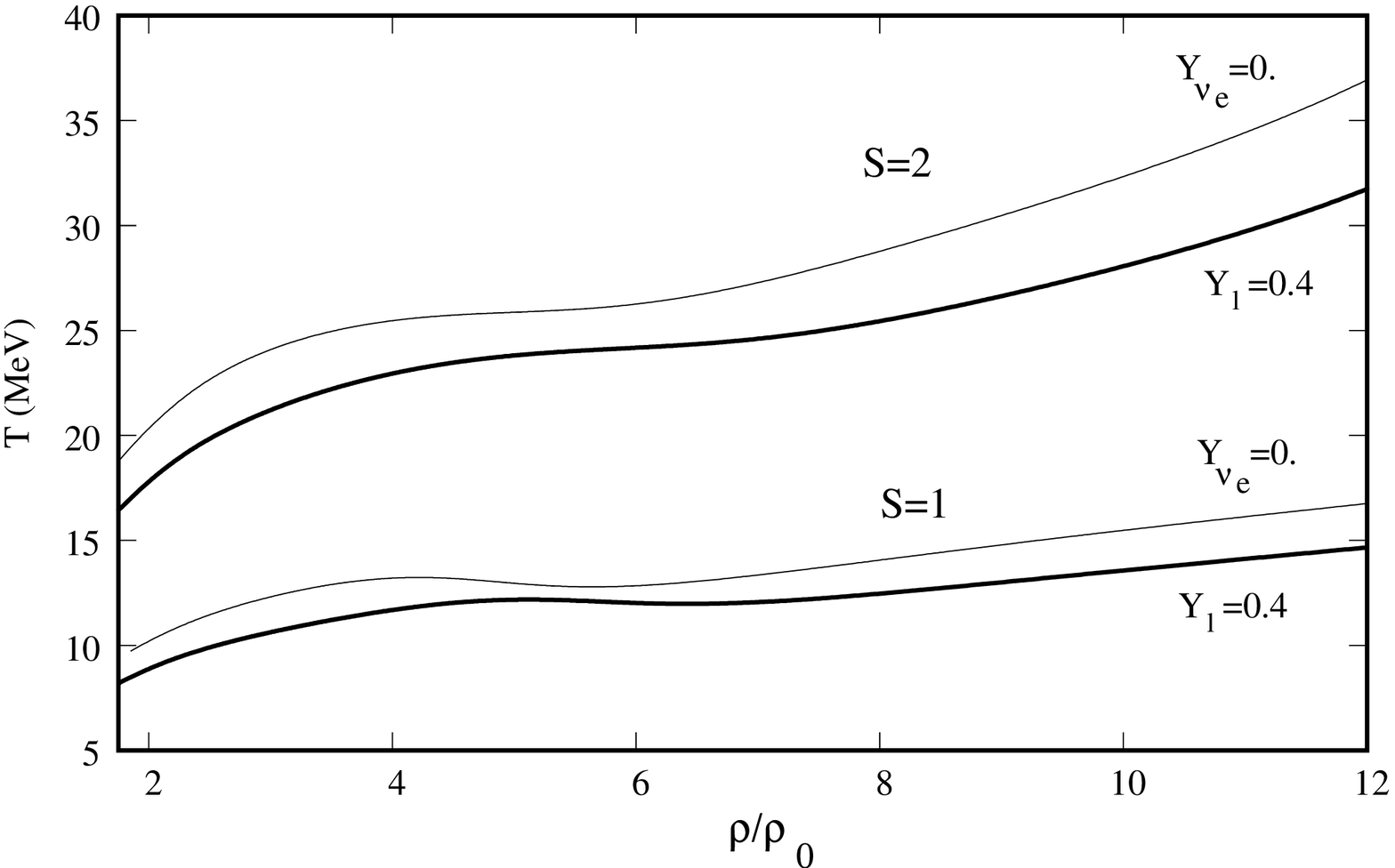}\\
\end{tabular}
\caption{Temperature range obtained for $S=1,2$ without neutrinos and with 
trapped neutrinos within the MIT bag model (top figure) and the NJL model 
(bottom figure).}
\label{fig2}
\end{figure}

In figure \ref{fig2} we show the temperature range described by both 
models for the cases $S=1$ and $S=2$. For the MIT model, the 
temperature for each different EOS increases steadily, reaching about 
$35\rm\,MeV$ for $S=2$ and $\simeq 17\rm\,MeV$ for $S=1$ at 
$\rho=12\rho_0$. The NJL model presents a different  behavior: the 
temperature first oscillates before increasing. The near-plateau 
around 4--6 $\rho/\rho_0$ coincides with the onset of strangeness 
opening of a new degree of freedom. The maximum temperature for $S=2$ 
without neutrinos is higher than the one obtained with the MIT model 
for the same energy density. This difference is due to the fact that 
within the MIT model the fractions of the different types of quarks 
are similar, unlike the NJL which has a smaller fraction of 
$s$-quarks, which gives rise to a higher entropy. If we fix the 
entropy the temperature must be lower within the MIT model.
However, from Table I we can see that within the NJL the central 
energy densities of compact stars are much lower than the corresponding ones 
for the MIT model. The temperatures are not shown in Table I because they vary
though the star; the temperatures in the interior of the quark stars are then
similar for the NJL and the MIT models, despite the significant difference in 
density. Including trapped neutrinos lowers the temperature of the star.
This is due to the presence of a fixed fraction of leptons 
increasing the degrees of freedom, and then to keep the 
entropy per particle and the thermodynamic potential fixed, 
a lower temperature is required. If neutrino trapping is 
not imposed the fraction of leptons in the star is very small.

\begin{figure}
\begin{tabular}{cc}
\includegraphics[width=10.cm]{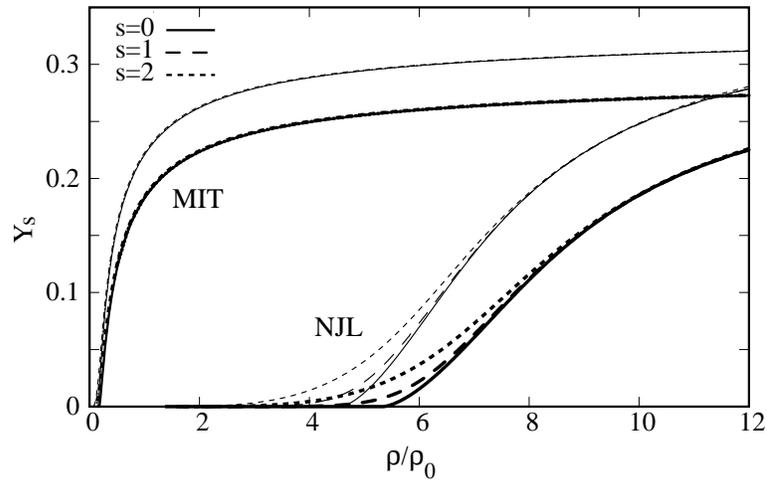}\\
\end{tabular}
\caption{Strangeness content obtained for $S=0,1,2$ without neutrinos and 
with trapped neutrinos within the MIT bag model (top figure) and the NJL model 
(bottom figure).} 
\label{fig3}
\end{figure}

In figure \ref{fig3} we plot the strangeness content in each 
situation described in the text. As already referred, within the NJL 
the onset of strangeness occurs for $\rho/\rho_0\sim 4$ depending on 
the temperature and the neutrino content. A different situation 
applies for the MIT bag: strangeness is present from lower densities 
and at $\rho/\rho_0\sim 4$ the strangeness fraction  is almost at its 
maximum value in the case without neutrinos, $\sim 0.33\,$. The 
presence of leptons in the case of trapped neutrinos with a fixed 
lepton fraction lowers the strangeness content due to the electric 
charge conservation restrictions.

\begin{figure}
\begin{tabular}{cc}
\includegraphics[width=10.cm]{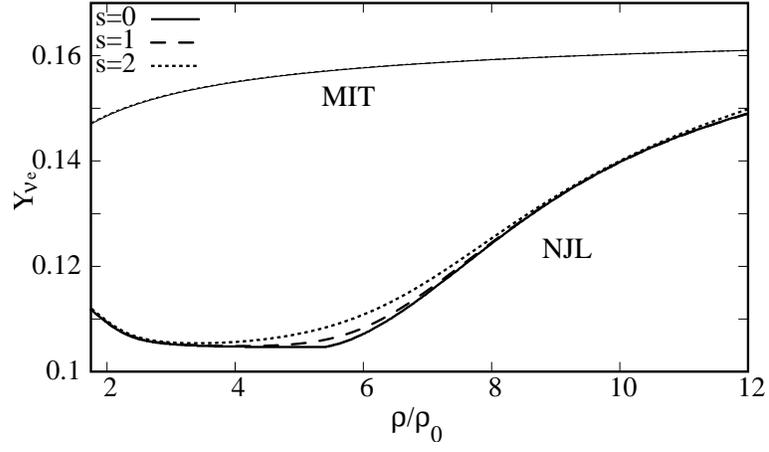} \\
\end{tabular}
\caption{Neutrino fraction obtained for $S=0,1,2$ within the MIT bag model 
(top curves) and the NJL model (bottom curves).} 
\label{fig4}
\end{figure}

In figure \ref{fig4} we display, for both the MIT and the NJL models, the 
electron neutrino fraction 
present in the interior of the stars when trapped neutrinos are included. 
With the MIT model the neutrino content is
practically the same for the three entropies considered. It increases 
a little for lower densities and then reaches its maximum value, for 
the same reasons discussed above, i.e., neutrinos and electrons 
together are fixed to a certain fraction and electrons compensate for 
charge neutrality in a system where the three quarks have
equal number densities. The NJL model shows a slight decrease in the 
neutrino fraction until the onset of
strangeness. This effect is more pronounced for $\rho<3\rho_0$ when 
chiral symmetry restoration has still not occured for the $u$ and $d$ 
quarks. A slight increase in the neutrino fraction occurs  with the 
increase of the entropy, mainly at low densities. The neutrino 
fraction never reaches 14$\%$ within the NJL picture, but it amounts 
to more than 15$\%$ within the MIT model. We see a similar neutrino
fraction for these models in \cite{trapping,pmp2}. The difference in 
neutrino content is again due to the larger fraction of strange 
quarks in the MIT description and therefore a smaller number density 
of electrons required by the electric charge neutrality. A smaller fraction 
of electrons implies a larger fraction of neutrinos for a fixed 
lepton fraction. Nevertheless, if the CFL model were used for the 
quark matter with trapping, a much higher neutrino fraction, of the 
order of 35$\%$, would be possible \cite{pmp2}.

Given the EOS, the next step is to solve the 
Tolman-Oppenheimer-Volkoff equations \cite{tov}. Table~I lists the 
main properties of the stars, including their maximum gravitational 
and baryonic masses, their radii and central energy densities, for 
all the EOSs considered in the present work. As expected, quark stars 
have smaller maximum  masses with smaller radii 
\cite{recentours,pmp1,pmp2} and higher central
densities.

Let us first analyze the results obtained with the MIT model and then 
outline the main differences that appear with the NJL model. For 
quark stars either without or with neutrinos, the maximum 
gravitational masses and radii are always around 1.23--1.24 
$M_{\odot}$ and 6.7--$7.1\rm\,$, respectively, and the
energy density is around $14.5\rm\,fm^{-4}$. The inclusion of trapped 
neutrinos and a crust, have practically no effect on the numerical 
results.

The NJL model leads to significantly different results. The maximum 
masses and radii are systematically larger and, as a consequence, the 
central energy  densities are lower than for the MIT bag model. 
In fact, the results do not depend strongly on
the entropy, as already pointed out in \cite{recentours}. 
In the NJL, for a bare star without neutrinos, the maximum masses and 
radii decrease and the central energy density increases as the 
entropy increases. For a star with a crust, the gravitational and
baryonic masses decrease but the radii oscillate as the entropy increases.

\begin{figure}
\begin{tabular}{cc}
\includegraphics[width=10.cm]{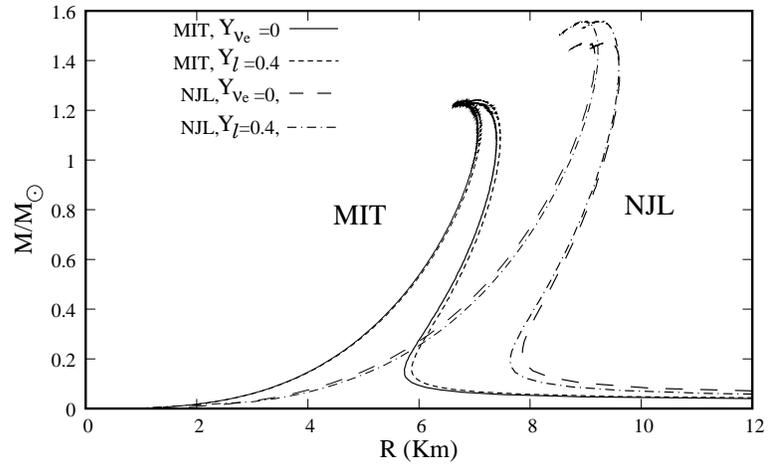}\\
\end{tabular}
\caption{Mass-radius plots for bare and crusted MIT and NJL stars at $S=0$ 
with and without neutrinos.}
\label{fig5}
\end{figure}

In Fig.~\ref{fig5} we show for $S=0$, and for both models the 
mass-radius plot of the family of stars obtained for the neutrino 
free and the  neutrino trapped situations. It is clear that the NJL 
predicts the possibility of existence of more massive and with larger 
radius stars. For a fixed radius smaller than the larger accepted by the MIT 
model the NJL gives stars with much smaller gravitation masses.
Conclusions for finite entropies are the same.

Another difference is that when trapped neutrinos are included, this 
generally increases the maximum gravitational mass, with the increase 
being much larger for the NJL than for the MIT model.

In a previous work \cite{recentours} we see that the MIT results are 
sensitive to the $Bag$ parameter. With a lower value of this 
parameter,results similar to the ones obtained with the NJL model are 
found.
The same is possible when the CFL model is used. In \cite{recentours} 
we fixed the gap parameter of the CFL model as $100\rm\,MeV$ and 
showed that the properties of the star depend on the $Bag$ parameter, 
but the properties are also known to depend on the gap parameter 
\cite{horvath}.

It is not clear that quark stars really exist, and if the do exist in 
principle, how they would be formed. They form directly in a 
supernova explosion, or as a consequence of the hadron-quark 
deconfinement phase transition in stellar compact stars 
\cite{bombaci}. In the former case there is neutrino trapping during 
the first seconds of the stars life and the entropy is approximatly 
constant throughout the star \cite{burrows}.  In this case one 
concludes from our results that there can be no late blackhole 
formation. This event is only possible if the baryonic mass of a stable
star at finite temperature is larger than the largest baryonic mass of a stable
cold star \cite{prak97}. From Table I one can see that the maximum baryonic 
masses at fixed entropies are always lower than the corresponding values at 
zero temperature. This kind of effect could occur within hybrid stars 
\cite{prak97,trapping}.

\begin{table}[h]
\begin{center}
\caption{Quark star properties for the EOSs described in the text.}
\begin{tabular}{lccccccc}
\hline
type & crust & entropy & neutrinos & $M_{\max}$ & $M_{b\max}$ & $R$ 
& $\varepsilon_0$\\
&&&& $(M_{\odot}$)&($M_{\odot}$)& (km) & (fm$^{-4}$)\\
\hline
MIT & no  & 0 & no  & 1.22 & 1.29 & 6.77 & 14.53 \\
MIT & no  & 1 & no  & 1.23 & 1.28 & 6.78 & 14.35 \\
MIT & no  & 2 & no  & 1.23 & 1.26 & 6.77 & 14.73 \\
MIT & yes & 0 & no  & 1.23 & 1.21 & 7.10 & 14.61 \\
MIT & yes & 1 & no  & 1.23 & 1.20 & 7.06 & 14.67 \\
MIT & yes & 2 & no  & 1.23 & 1.18 & 7.01 & 14.54 \\
\hline
MIT & no  & 0 & yes & 1.24 & 1.19 & 6.82 & 14.36 \\
MIT & no  & 1 & yes & 1.23 & 1.18 & 6.81 & 14.52 \\
MIT & no  & 2 & yes & 1.24 & 1.18 & 6.86 & 13.98 \\
MIT & yes & 0 & yes & 1.24 & 1.03 & 7.10 & 14.41 \\
MIT & yes & 1 & yes & 1.24 & 1.02 & 7.11 & 14.41 \\
MIT & yes & 2 & yes & 1.24 & 1.00 & 7.11 & 14.41 \\
\hline
NJL & no  & 0 &  no & 1.47 & 1.56 & 9.03 &  7.26 \\
NJL & no  & 1 &  no & 1.47 & 1.54 & 9.02 &  7.46 \\
NJL & no  & 2 &  no & 1.46 & 1.51 & 9.01 &  7.60 \\
NJL & yes & 0 &  no & 1.47 & 1.55 & 9.34 &  7.44 \\
NJL & yes & 1 &  no & 1.47 & 1.53 & 9.35 &  7.52 \\
NJL & yes & 2 &  no & 1.46 & 1.50 & 9.29 &  7.76 \\
\hline
NJL & no  & 0 & yes & 1.56 & 1.57 & 8.99 &  8.07 \\
NJL & no  & 1 & yes & 1.55 & 1.55 & 8.99 &  8.04 \\
NJL & no  & 2 & yes & 1.54 & 1.53 & 9.00 &  8.07 \\
NJL & yes & 0 & yes & 1.56 & 1.56 & 9.26 &  8.15 \\
NJL & yes & 1 & yes & 1.55 & 1.55 & 9.23 &  8.19 \\
NJL & yes & 2 & yes & 1.54 & 1.52 & 9.25 &  8.23 \\
\hline
\end{tabular}
\end{center}
\end{table}

From figure \ref{fig5} and Table I we can also see that the inclusion of 
a crust has a similar effect for both the MIT and NJL models: the radius of 
the maximum mass star increases $\sim0.3\rm\,km$. It has been argued that 
at finite temperature the star tends to be bare due to the reduction 
of the electrostatic potential of the electron \cite{glen95}. In 
particular, at $T=30\rm\,MeV$ a star has essentially no crust. Taking 
into account simulations with neutron stars, we
should  consider the  entropy rather than the  temperature to be 
constant throughout the star \cite{burrows}. In this case the 
temperature at the surface is lower: we get $\sim8$ and $16\rm\,MeV$
respectively for $S=1$ and $S=2$, respectively.

\begin{figure}
\begin{tabular}{cc}
\includegraphics[width=8.cm]{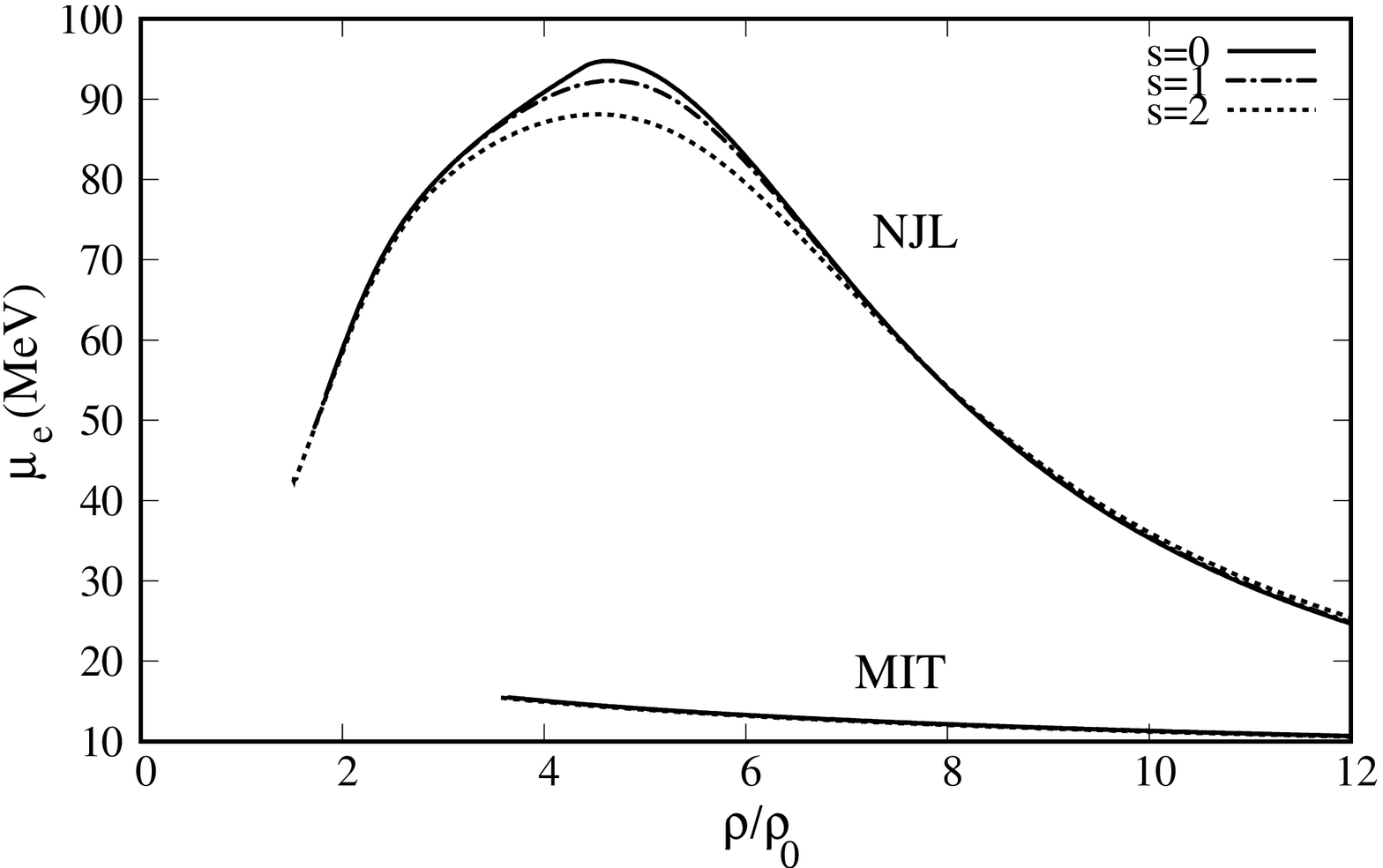}\\
\includegraphics[width=8.cm]{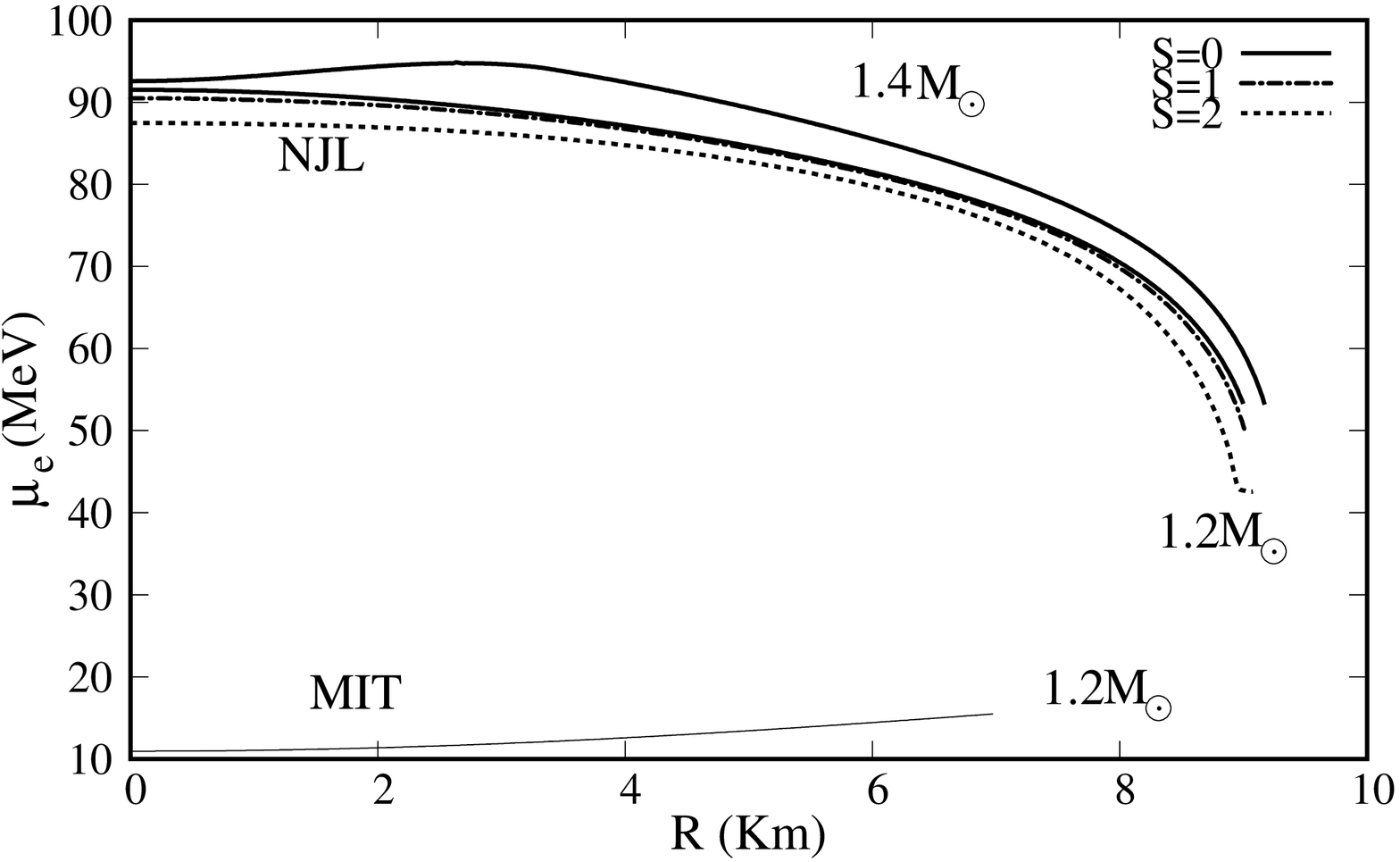} \\
\end{tabular}
\caption{The electron chemical potential a) as a function of density; 
b) in the interior of stars with $M=1.2$ and 1.4 $M_\odot$. Thin lines are 
for the MIT model and thick lines for the NJL.} 
\label{mue}
\end{figure}

An important difference between the NJL stars and MIT stars is that 
the electron chemical potential is very different. In 
Fig.~\ref{mue}a  we show, as a function of the density, the electron 
chemical potential in both models for the different values of 
entropy. Entropy has almost no effect on the results but there is an 
important difference between the MIT and the NJL cases: $\mu_e < 20 
\rm\,MeV$ for the MIT model and $\mu_e$ as high as $100\rm\,MeV$ for 
the NJL model. In the interior of a star the electron chemical potential
is shown to be equal to the electric potential energy 
\cite{glen95,chmaj}. In  figure \ref{mue}b we plot this quantity for stars 
with $M=1.2 M_\odot$ for NJL (thick lines) and MIT (thin lines), and  for 
stars with $M=1.4 M_\odot$ for NJL (thick lines). In the last two cases, 
which correspond to stars close to the maximum mass for a stable 
star, the curves for  $S=0,1$ and 2 are almost coincident and so we 
have only plotted the $S=0$ curve.

One of the motivations of this work was to verify  whether, with the 
NJL model, the presence or absence of a crust on a quark star affects 
its properties and how these properties differ from those obtained 
with the MIT model. We already knew that within the MIT and the CFL 
models the differences are non-negligible \cite{recentours}. From 
Table~I and Fig.~\ref{fig5} the presence of the crust, described by 
the BPS EOS \cite{bps}, affects the MIT model than the NJL model. 
With the NJL model  the main differences appear in the radius and the central 
energy density.

We note that our results are different from those obtained in 
\cite{hanauske}, where a lower maximum mass was obtained, probably 
due to a different choice of the parametrization for the NJL model.

\section{Conclusions}

In this paper we consider the properties of quark stars with the NJL 
model for the EOS.  This model has more realistic features than the 
MIT bag model, in particular, chiral symmetry. Due to the large 
$s$-quark mass at low densities, several features of the NJL stars 
are different from the corresponding quantities for the MIT stars: 
smaller  strangeness content, higher electron chemical potentials, 
smaller neutrino
fractions in stars with trapped neutrinos, higher maximum star masses 
and  smaller maximum central densities. In fact, the properties of 
quark stars with the NJL model are closer to the corresponding 
properties of hybrid stars, e.g. maximum masses, than to the 
properties of the MIT quark stars.

We also show that if a quark star is formed directly from a supernova 
explosion there can be no delayed blackhole formation, because the 
maximum baryonic masses of hot stars are always smaller that the 
corresponding masses of cold stars.

The electron chemical potential inside the stars has quite large values  
with the NJL model. This influences the properties of the surface of a 
quark star and the possibility of crust formation.

\section*{Acknowledgements}

This work was partially supported by Capes (Brazil) under process
BEX 1681/04-4, CAPES (Brazil)/GRICES (Portugal) under project 100/03 and
FEDER/FCT (Portugal) under  the project  POCTI/FP/FNU/50326/2003.
  D.P.M. would like to thank the friendly atmosphere at the
Reserch Centre for Theoretical Astrophysics, Sydney University, where
this work was done.

\end{document}